# Experimental evidence for electron-electron interaction and spin-charge separation in graphene quantum dots

Hui-Ying Ren[§], Ya-Ning Ren[§], Qi Zheng[§], Jia-Qi He, and Lin He*

**Affiliations:**
Center for Advanced Quantum Studies, Department of Physics, Beijing Normal University, Beijing, 100875, People's Republic of China

[§]These authors contributed equally to this work.
[†]Correspondence and requests for materials should be addressed to Lin He (e-mail: helin@bnu.edu.cn).

**Graphene quantum dots (GQDs) can exhibit a range of spectacular phenomena such as the Klein-tunneling-induced quasibound states[1-6] and Berry-phase-tuned energy spectra[7-15]. According to previous studies, all these interesting quantum phenomena seem to be well understood in the free electron picture[1-15]. However, electronic motion in the GQDs is reduced to quantized orbits by quantum confinement, which implies that the kinetic energy may be comparable to or even smaller than the Coulomb energy of the quasiparticles, possibly resulting in exotic correlated phases in the GQDs. Here we present a scanning tunneling microscopy and spectroscopy study of gate-tunable GQDs in graphene/WSe$_2$ heterostructure devices and report for the first time the observation of electron-electron interaction and correlation-induced spin-charge separation in the GQDs. Gating allows us to precise characterize effects of the electron-electron interaction on the energy spectra of the GQDs. By measuring density of states as a function of energy and position, we explicitly uncover two density waves with different velocities in the GQDs, attributing to spin-charge separation in real space.**

Exotic many-body quantum phases often occur in systems with strong electron-electron (e-e) interactions[16-20]. Creating a flat band is a well-established method to realize strongly correlated systems, in which the e-e interactions $U$ can largely exceed the kinetic energy of electrons set by the bandwidth $W$. Therefore, significant efforts have been devoted to the search for emergent quantum phenomena in Landau levels and moiré flat bands. These studies have achieved great success and many exotic correlated phases, such as the fractional quantum Hall effect, superconductivity, and quantum anomalous Hall effect, have been observed[16-20].

Recent advances in introducing quantum confinement in a continuous graphene system have provided a new route to strongly suppress the kinetic energy of quasiparticles[1-15,21-25]. In graphene quantum dots (GQDs), electronic motion is locally reduced to quantized quasibound states by quantum confinement, *i.e.*, the Klein-tunneling-induced whispering gallery mode (WGM)[1-6], implying that the e-e interactions should play an important role in determining their electronic properties. However, almost all the interesting quantum phenomena reported in the GQDs can be well understood in the free electron picture[1-15,21-25]. Although some spectroscopic evidence of e-e interactions in the GQDs has been observed, clear identification of e-e interactions on the energy spectra is still missing due to the metallic substrate and the lack of gate control[26]. Moreover, no correlated state has been reported in the GQDs up to now. Here we report for the first time the observation of e-e interactions and correlation-induced spin-charge separation in the GQDs embedded in gate-tunable graphene/$WSe_2$ heterostructure devices. Our gate-tunable devices enable unambiguous measurement of effects of the e-e interactions on the energy spectra of the GQDs, which are consistent with expectations for a particle in a box with interaction-induced level splitting. The e-e interactions and the suppressed quantum kinetic energy lead to a new correlated phase in the GQDs. Two density waves with different velocities, attributing to spin-charge separation, are explicitly uncovered in the GQDs in real space through measurement of density of states (DOS) as a function of energy and position.

Figure 1a shows schematic of the experimental device set-up. In our experiment, the thick $WSe_2$ layers are separated from the bulk crystal by traditional mechanical exfoliation technology and are transferred to the $Si/SiO_2$ substrate with Au electrodes. Then the graphene monolayer is put on the $WSe_2$ layers by wet transfer technology to obtain the graphene/$WSe_2$ heterostructure. A voltage pulse from scanning tunneling microscope (STM) tip is used to create monolayer $WSe_2$ QD at the interface of the graphene/$WSe_2$ heterostructure (a nanoscale pit is generated simultaneously in the $WSe_2$ substrate, indicating that the $WSe_2$ QD is created from the pit by the tip pulse)[6,27]. A STM image of a representative graphene/$WSe_2$ heterostructure QD is shown in Fig. 1b (see Fig. S1 for more experimental data). The lower panel shows a profile line across the QD, indicating that the thickness is the same as that of a $WSe_2$ monolayer ~ 0.8 nm. The $WSe_2$ QDs can introduce electrostatic potentials on the graphene above them, which generate GQDs to confine massless Dirac fermions of graphene in the graphene/$WSe_2$ heterostructure[6,27]. Figure 1c shows a typical scanning tunneling spectroscopy (STS), *i.e.*, d$I$/d$V$, spectroscopic map across the GQD and a sequence of resonance peaks, arising from temporarily-confined quasibound states, can be clearly observed. At the edge of a GQD, the quasibound states are generated by the Klein-tunneling-induced WGM and they are almost equally spaced in energy (see left panel of Fig. 1c for a representative STS spectrum)[1-8]. For a circular GQD with the size of confined region $L$, the energy spacing $E_0$ between the quasibound states should be described by $\Delta E \approx 2\hbar v_F/L$, where $\hbar$ is the reduced Planck's constant and $v_F$ is the Fermi velocity. In Fig. 1d, we summarize the energy spacing for the GQDs with different sizes observed in our experiment. Fitting the data yields the Fermi velocity $v_F \approx 1.2 \times 10^6$ m/s, further confirming the confinement of the massless Dirac fermions in the GQDs. The deviation of the experimental data from the fitted line mainly arises from the deviation of the structures of the studied GQDs from the ideal circular GQD (see Fig. S1).

A notable feature observed in our experimental is that the energy spacing between the two quasibound states flanking the Fermi level (i.e., the highest occupied and the lowest unoccupied quasibound states), labelled as $E_{large}$, is much larger than the $E_0$ ($E_0$ is the energy spacing between the two quasibound states that both are at one side of the Fermi level), as shown in Fig. 1c (see Fig. S2 for more experimental results obtained on different GQDs). Similar results have been observed in the two flat bands of magic-angle twisted bilayer graphene (MATBG)[28-31] and in quantized confined states of finite-size Tomonaga-Luttinger liquids (TLLs)[32]. Both the energy separation between the two flat bands in the MATBG and the energy difference between the two confined states in the finite-size TLLs are much larger when they are flanking the Fermi level, which are clear signatures of the e-e interactions[28-32]. In the GQDs, the energy difference between the $E_{large}$ and the $E_0$ directly reflects the e-e interaction strength $E_C$, i.e., $E_{large} \approx E_0 + E_C$. The Coulomb energy $E_C$ of two electrons should be inversely proportional to their separation $L$, i.e., $E_C$ is expected to scale as $L^{-1}$. Therefore, the $E_{large}$ should also be inversely proportional to $L$, which is demonstrated explicitly in our experiment, as shown in Fig. 1d.

To further explore the e-e interactions on the energy spectra of the GQDs, we perform STS measurement of the GQDs for different gate voltages. Figure 2a shows the gate-dependent d$I$/d$V$ curves acquired at a GQD with $L \approx 19$ nm (see Supplemental materials Fig. S3 for STM characterizations of the GQD). A key observation is that the energy separation between the two peaks flanking the Fermi level changes with gate voltage. Two representative STS spectra at $V_g = 18$ V and $V_g = -4$ V are plotted in Fig. 2b. At $V_g = 18$ V, the energy separation between the $N_2$ and $N_3$ states, the $E_{large}$, is about 115 meV, which is larger than that ~ 79 meV between the $N_1$ ($N_3$) and $N_2$ ($N_4$) states, the $E_0$. For $V_g = -4$ V, the $N_3$ state is nearly half filled and is split into two peaks, the $N_{3+}$ and $N_{3-}$, with reduced intensities. Then the energy separation between the $N_{3+}$ and $N_{3-}$, i.e., the $E_C$, is about 35 meV. Such a measurement demonstrates explicitly that $E_{large} \approx E_0 + E_C$ (see Fig. S4 for more gate-dependent measurements on the other GQD and the same result can be obtained). The observed gate-induced modulation of the energy spectra reveals that e-e interaction plays an important role in determining electronic properties of the GQDs. When e-e interaction is absent, as schematically shown in Fig. 2c (left panel), the energy separation between the quasibound states would be gate and energy independent and should be equal to the single-particle level spacing value in a box. As schematically shown in Fig. 2c (right panel), the presence of e-e interaction explains well the large and small energy separations observed for different energies (results in Fig. 1) and different electron fillings (results in Fig. 2).

In the GQDs, the e-e interaction is about several tens millielectron volt and is comparable to the full width at half maximum of the quasibound states (results in Figs. 1 and 2), possibly causing correlation physics to manifest experimentally by the emergence of new quantum states. To further explore correlated phases in the GQDs, we measure d$I$/d$V$ spectra of a GQD and perform Fourier transform (FT) analysis of the resulting density plot, as summarized in Figs. 3a-3c. Here we choose such a GQD (Fig. 3a) because that the quasi-one-dimensional confinement along the long side much simplifies the analysis of real-space standing waves originating from the constructive or destructive interference of the confined quasiparticles (as illustrated in Fig. 4 subsequently). Moreover, a longer confinement along the long side of the GQD also allows us to achieve better momentum resolution. Figure 3b shows the STS intensity plot as a function of position ($x$ axis) along the arrow in Fig. 3a and the sample bias ($y$ axis), which directly reflects real-space modulation of the local DOS at different energies. The number of nodal points of the local DOS increases with changing the energies away from the Dirac point (at about 246 meV in the GQD). For example, the number of nodal points is 2 at the energy of 175 meV, and it increases to 5 at the energy of 30 meV. The corresponding FT of the STS intensity plot, as shown in Fig. 3c, directly reveals the dispersion of the quasibound states in the GQD. Unexpectedly, two linear dispersion branches with different velocities, $v_{F1} \approx 1.09 \times 10^6$ m/s and $v_{F2} \approx 1.36 \times 10^6$ m/s, are seen to cross the Dirac point (See Fig. S5 for more experimental results on different GQDs and see Supplemental materials for details of the fitting in determining the Fermi velocities). Such a result is quite different from the expected result in the free electron picture, i.e., a density wave with one velocity, as schematically shown in top panel of Fig. 3d.

The existence of two density waves with different velocities in the GQD is further confirmed by carrying out STS mappings, which directly reflect the local DOS in real space at the selected energies. In a quantum confined system with a fixed size, the two density waves with different velocities will form the standing waves at different discrete energies, which provide characteristic fingerprints in the STM measurement. Figure 4a shows representative STS maps measured at 5 different energies of the GQD in Fig. 3a. Figure 4b shows typical profile lines of the 5 STS maps and the values of the profile lines at the boundaries of the confined potential are shifted to zero for comparison. The corresponding pattern expected for free particles with $v_{F2} \approx 1.36 \times 10^6$ m/s and the Dirac point $E_D \approx 246$ meV confined to a 1D box of the same length is shown in Fig. 4c. Three significant discrepancies are present compared to Fig. 4c: In the experiment (1) the number of maxima in the DOS at 30 meV is four; (2) the minimum between two maxima in the DOS is much larger than zero at 115 meV and 30 meV; (3) the spatial separation between adjacent two maxima is not approximately equidistant at 30 meV. These discrepancies can be naturally explained by considering the existence of two density waves with different velocities and the correlation-induced large energy separation of the quasibound states around the Fermi level. Figure 4d shows the DOSs at different energies by superposition two density waves with $v_{F1} \approx 1.09 \times 10^6$ m/s, $v_{F2} \approx 1.36 \times 10^6$ m/s and the Dirac point $E_D \approx 246$ meV (determined experimentally in Fig. 3c), which reproduce well the main features observed in our experiment.

The emergence of correlation-induced two density waves with different velocities in the GQDs reminds us the spin-charge separation induced by e-e interactions in the TLLs[32-35]. The two dispersions are reasonably attributed to spin- and charge-density waves with different velocities. In different GQDs, the velocities for the spin- and charge-density waves are different (see Fig. 3 and Fig. S5), which is similar as that observed in the finite-size TLLs[32]. Here we should also point out that the spin-charge separation observed in the GQDs is quite different from that in the TLLs. First, the ratio between the velocities for the spin- and charge-density waves $v_{F1}/v_{F2}$ in the GQDs is about ~ 0.75±0.05, which is smaller than that, ~ 0.53±0.05, observed in the finite-size TLLs very recently[32], possibly due to the relative weaker e-e interactions in the GQDs. Second, as schematically sketched in Fig. 3d, the highest occupied and the lowest unoccupied states in the TLLs are the so-called zero modes, i.e., no spin or charge modes are excited[32-35]. Injecting electrons or holes into the TLLs will create spin and charge excitations with different velocities (Fig. 3d, right bottom panel). In the GQDs, the correlation induced spin- and charge-density waves with different velocities emerge for all the temporarily-confined quasiparticles, i.e., for all the quasiparticles below the Dirac point (Fig. 3d, left bottom panel). Therefore, a new theory with different mechanism of the TLLs should be constructed to fully understand the emergence of correlation-induced spin-charge separation in the GQDs.

In the experiment, the relative magnitude of the maxima in the local DOS, as shown in Fig. 4b, is not captured by the simulated result in Fig. 4d. This is reasonable because that the simulation is based on several simplified assumptions: (1) the confinement of the electronic potential on the two density waves is the same; (2) the two density waves contributed equally to the measured local DOS; (3) the damping coefficients for the two density waves are the same at positions that deviate from the formation of the standing waves in a box; (4) there is no interaction between the two density waves. All these factors can affect the relative magnitude of the standing waves observed in experiment. A notable difference between the results in Fig. 4b and Fig. 4d is that the observed local DOS at -50 meV in experiment seems to be only contributed from the charge density waves and there is almost no contribution from the spin density waves. In indicates that the electronic potential has negligible confinement on the spin density waves at this energy. Such a result is consistent with the FT result obtained in Fig. 3c, in which no signal from the spin density waves is detected at -50 meV.

In summary, the effects of the e-e interactions on the energy spectra of the GQDs are unambiguously measured in our gate-tunable devices. In the GQDs, two density waves with different velocities, attributing to spin-charge separation in real space, are explicitly uncovered. These interesting phenomena cannot be explained in the absence of e-e interactions and highlight the importance of correlations in the GQDs. Our result suggests that the spin-charge separation may be a universal ground state in finite-size systems, such as TLLs and GQDs, with strong correlated effects. Further theoretical and experimental work are necessary to fully ascertain the importance of correlation effects in the GQDs.

**Acknowledgments**

This work was supported by the National Key R and D Program of China (Grant Nos. 2021YFA1400100 and 2021YFA1401900) and National Natural Science Foundation of China (Grant Nos. 12141401, 11974050).

**Author contributions**

**Methods**

**CVD Growth of Graphene.** The large area graphene monolayer films were grown on a $20 \times 20$ mm$^2$ polycrystalline copper (Cu) foil (Alfa Aesar, 99.8% purity, 25 µm thick) via a low-pressure chemical vapor deposition (LPCVD) method. The cleaned Cu foil was loaded into one quartz boat in center of the tube furnace. Ar flow of 50 sccm (Standard Cubic Centimeter per Minutes) and H$_2$ flow of 50 sccm were maintained throughout the whole growth process. The Cu foil was heated from room temperature to 1030 ºC in 30 min and annealed at 1030 ºC for six hours. Then CH$_4$ flow of 5 sccm was introduced for 20 min to grow high-quality large area graphene monolayer. Finally, the furnace was cooled down naturally to room temperature.

**Construction of graphene/WSe$_2$ heterostructure.** We used conventional wet etching technique with polymethyl methacrylate (PMMA) to transfer graphene monolayer onto the substrate. PMMA was first uniformly coated on Cu foil with graphene monolayer. We transferred the Cu/graphene/PMMA film into ammonium persulfate solution, and then the underlying Cu foil was etched away. The graphene/PMMA film was cleaned in deionized water for hours. The WSe$_2$ crystal was separated into thick-layer WSe$_2$ sheets by traditional mechanical exfoliation technology and then transferred to the Si/SiO$_2$ substrate with Au electrodes. Then the graphene/PMMA film is put onto the WSe$_2$, ensuring that the graphene contacts with the Au electrode. Lastly, after the film is dried, the PMMA is removed using acetone and the sample is cleaned by alcohol to get the graphene/WSe$_2$ devices.

**STM and STS Measurements.** STM/STS measurements were performed in low-temperature (77 K and 4.2 K) and ultrahigh-vacuum (~$10^{-10}$ Torr) scanning probe microscopes [USM-1400 (77 K) and USM-1300 (4.2 K)] from UNISOKU. The tips were obtained by chemical etching from a tungsten wire to minimize tip-induced band bending effects of graphene. The differential conductance ($dI/dV$) measurements were taken by a standard lock-in technique with an ac bias modulation of 5 mV and 793 Hz signal added to the tunneling bias.

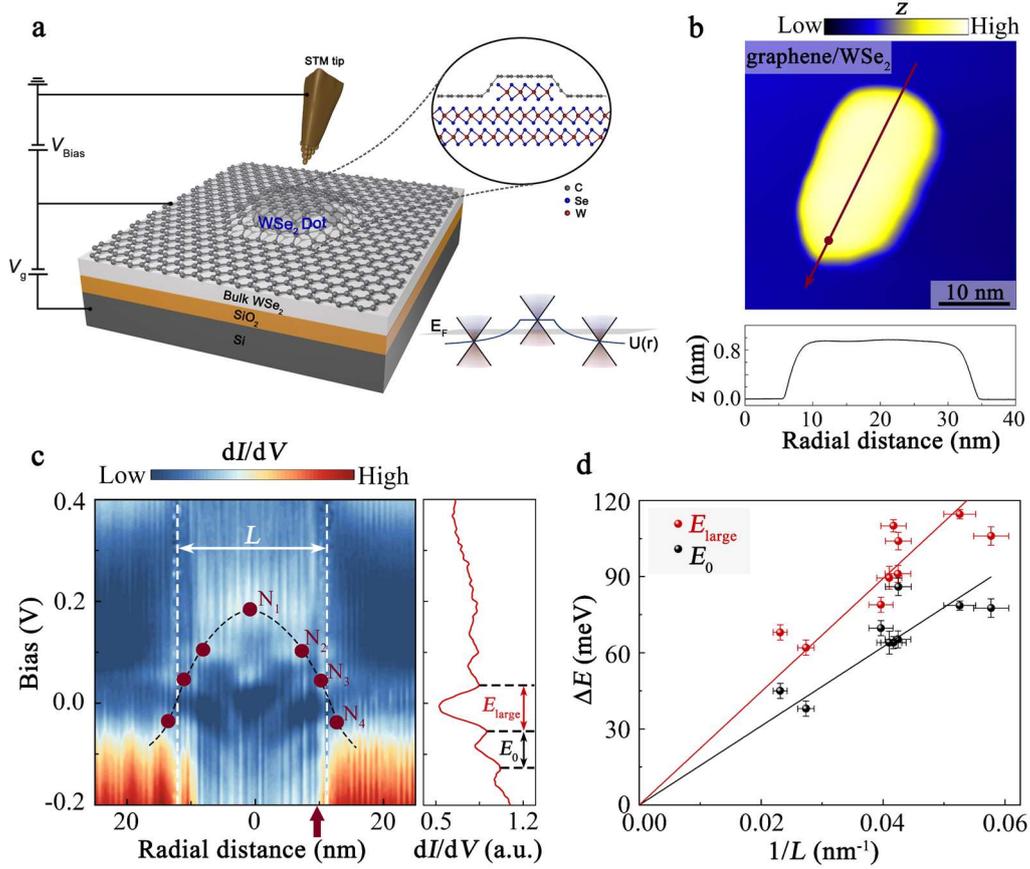

**Fig. 1. STM characterization of GQDs. a**, Schematic of the experimental device set-up. $V_b$ is the bias voltage applied between the sample and STM tip. The electron density is tuned by a Si gate beneath the bottom $SiO_2$ layer. **b**, Top panel: STM topography of a typical GQD (set-point parameters: $V_b$ = 600 mV, $I_{set}$ = 100 pA). Bottom panel: Profile line chart along the red straight line with an arrow in top panel shows the height of the GQD ~ 0.8 nm. **c**, Left panel: A radially d$I$/d$V$ spectroscopic map along the arrow in panel **b**. The red solid dots indicate the quasibound states, and the two white dashed lines mark the size of confined region $L$. Right panel: A d$I$/d$V$ spectrum recorded at the position indicated by the red solid dot in panel **b**. **d**, Measured large (red) and normal (black) energy separations between the quasibound states for the GQDs of different sizes. Both of them scale as $1/L$. The straight lines represent linear fits to the data.

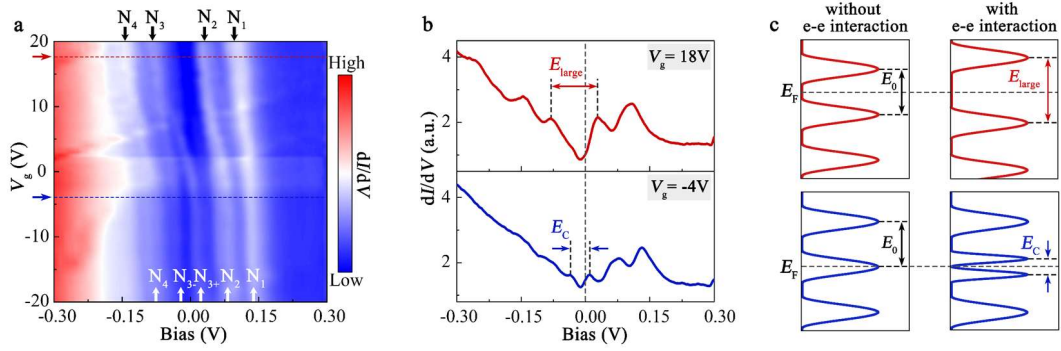

**Fig. 2. Gate-dependent electronic structure of the GQD. a,** A density plot of d$I$/d$V$ spectra acquired for -20 V ≤ $V_g$ ≤ 20 V on the GQD showing the energy separation transition as a function of $V_g$ (set-point parameters: $V_b$ = 700 mV, $I_{set}$ = 300 pA). The black or white arrows indicate the resonance peaks arising from the quasibound states. **b,** Two typical d$I$/d$V$ spectra at $V_g$ = 18 V and -4 V respectively, which are marked with the arrows in **a**. The $N_3$ state splits into two peaks with lower intensities when it is near half filled. **c,** Top panels: Quasibound-state diagram for the large energy separation case with (right panel) and without (left panel) e-e interactions. Bottom panels: Quasibound-state diagram for the small energy separation case with (right panel) and without (left panel) e-e interactions.

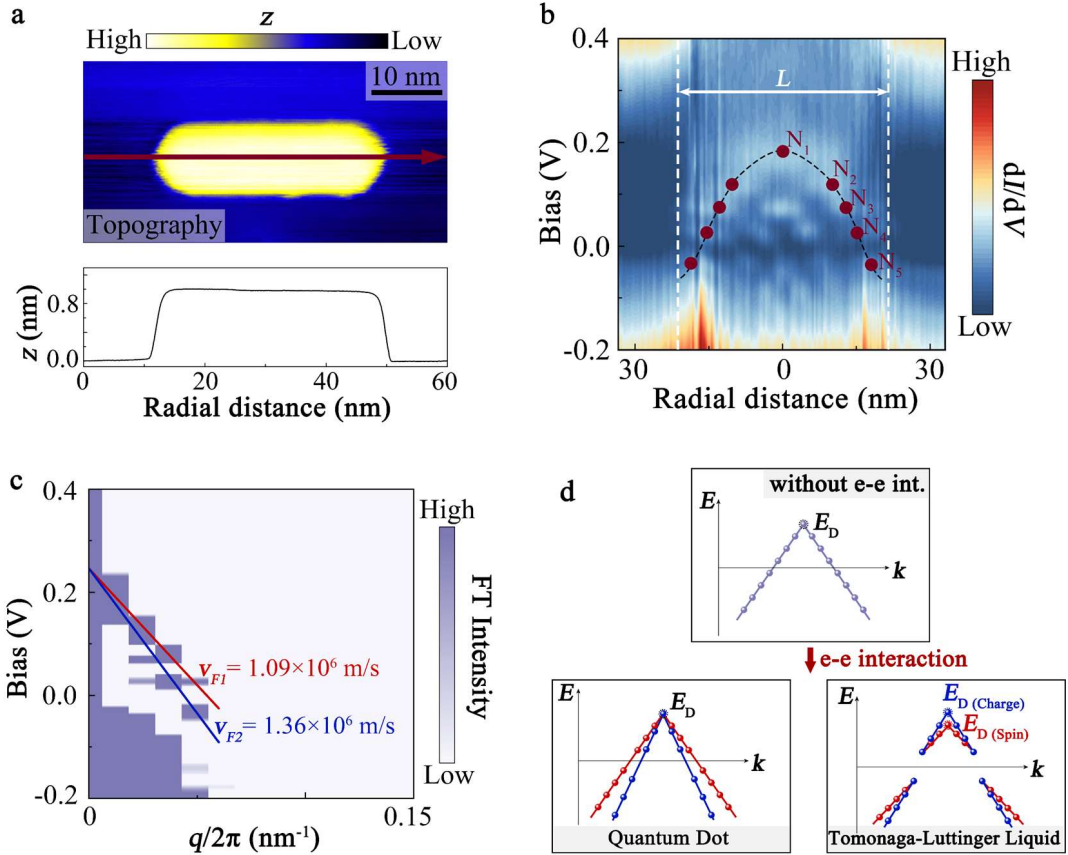

**Fig. 3. Dispersion of quasibound states in a GQD. a,** Top panel: STM topography of a typical GQD (set-point parameters: $V_b$ = 600 mV, $I_{set}$ = 100 pA). Bottom panel: Profile line chart along the red straight line with an arrow in top panel. **b,** A radially d$I$/d$V$ spectroscopic map along the red straight line with an arrow in **a**. The red solid dots indicate the quasibound states, and the two white dashed lines mark the size of confined region. The position of the Dirac point $E_D$ is ~ 246 meV and the size of confined region $L$ is ~ 43 nm. **c,** FT of the d$I$/d$V$ data in **b** as a function of bias voltage $V_b$ and wavevector $q$. Two linear dispersion branches with different velocities (marked red and blue) are observed. **d,** Schematic diagram of dispersion in a one-dimensional GQD. Top panel: dispersion in the GQD without e-e interactions. Bottom left panel: dispersion in the GQD with e-e interactions. There are two density waves with different velocities. Bottom right panel: spin-charge separation in the Tomonaga-Luttinger liquids.

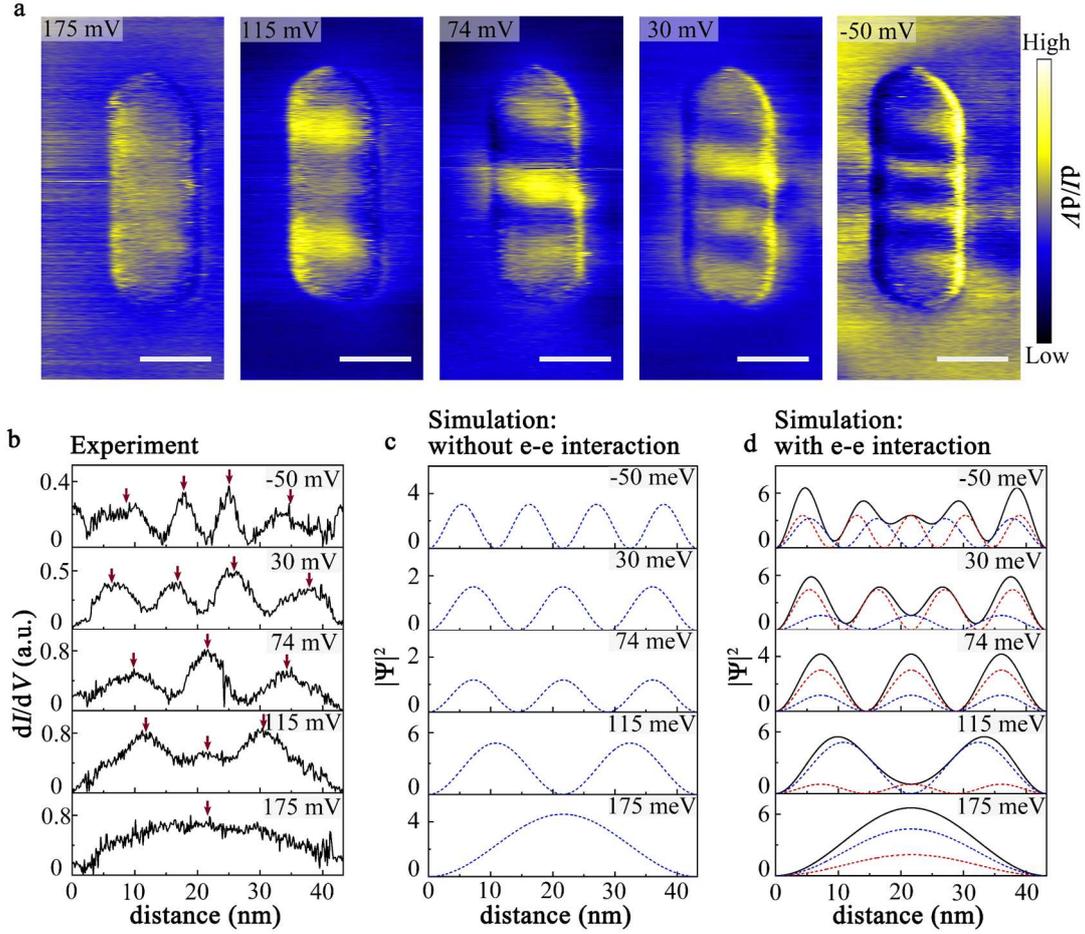

**Fig. 4. Imaging two density waves with different velocities in real space. a,** Representative STS maps recorded at different energies of the GQD in Figure 3a. Standing waves with different number of maxima in the DOS are formed at different energies due to the quantum confinement. **b,** Typical profile lines of the 5 STS maps in **a** with the size of confined region $L \approx 43.3$ nm. **c,** Theoretical local DOS assuming noninteracting quasiparticles with the Fermi velocity $1.36 \times 10^6$ m/s and the Dirac point $E_D = 246$ meV confined in a one-dimensional box with $L \approx 43.3$ nm. **d,** Theoretical LDOS by considering the confinement of two density waves with different velocities, $v_{F1} = 1.09 \times 10^6$ m/s, $v_{F2} = 1.36 \times 10^6$ m/s, and the Dirac point $E_D = 246$ meV confined in a one-dimensional box with $L \approx 43.3$ nm. The correlation-induced large energy separation of the quasibound states around the Fermi level is considered in the simulation. Scale bar: 10 nm in **a**.